\documentclass[twoside,draft]{article}

\usepackage[letterpaper]{geometry}
\usepackage[utf8]{inputenc}
\usepackage[english]{babel}

\usepackage{amssymb,amsfonts,amsmath}

\usepackage[perpage,symbol*]{footmisc}
\usepackage[final]{graphicx}
\usepackage{pstricks}
\usepackage{cite}

\usepackage[varg]{txfonts}

\usepackage{bm}

\begin{document}

\renewcommand{\refname}{References}
\renewcommand{\tablename}{\small Table}
\renewcommand{\figurename}{\small Fig.}
\renewcommand{\contentsname}{Contents}

\twocolumn[%
\begin{center}
\renewcommand{\baselinestretch}{0.93}
{\Large\bfseries Non-commutativity: Unusual View
}\par
\renewcommand{\baselinestretch}{1.0}
\bigskip
Valeriy V. Dvoeglazov\\ 
{\footnotesize  UAF, Universidad Aut\'onoma de Zacatecas
Apartado Postal 636, Suc. 3, C. P. 98061, Zacatecas, Zac., M\'exico.\rule{0pt}{12pt}
E-mail: valeri@fisica.uaz.edu.mx\\
}\par
\medskip
{\small\parbox{11cm}{%
Some ambiguities have recently been found in the definition of the partial
derivative (in the case of presence of both explicit and implicit
dependencies of the function subjected to differentiation). 
We investigate the possible influence of this subject on quantum
mechanics and the classical/quantum field theory. Surprisingly, some
commutators of operators of space-time 4-coordinates and those of 4-momenta 
are {\it not} equal to zero. We postulate the non-commutativity of 4-momenta and we derive 
mass splitting in the Dirac equation.
Moreover, two iterated limits may not commute each other, in general. Thus, we present an example 
when the massless limit of the function of $E, {\bf p}, m$ does not exist in some calculations 
within quantum field theory.
}}\smallskip
\end{center}]{%

\markboth{Valeriy Dvoeglazov. Non-commutativity: Unusual View.}{\thepage}
\markright{Valeriy Dvoeglazov. Non-commutativity: Unusual View.}

\section{Introduction.}

The assumption that the operators of
coordinates do {\it not} commute $[\hat{x}_{\mu },\hat{x}_{\nu }]_{-}\neq 0$
has been made by H. Snyder~\cite{snyder}. Therefore, the Lorentz symmetry may be
broken. This idea~\cite
{noncom,kruglov} received attention in the context of ``brane theories''.
Moreover, the famous Feynman-Dyson proof of Maxwell equations~\cite{FD}
contains intrinsically the non-commutativity of velocities $[\dot x_i (t), 
\dot x_j (t)]_- \neq 0$ that also may be considered as a contradiction with
the well-accepted theories (while there is no any contradiction therein).

On the other hand, it was recently discovered that the concept of partial
derivative is {\it not} well defined in the case of both explicit and
implicit dependence of the corresponding function, which the derivatives act
upon~\cite{brown}.
The
well-known example of such a situation is the field of an accelerated
charge~ \cite{landau}.\footnote{%
Firstly, Landau and Lifshitz wrote that the functions depended on $t^{\prime }$,
and only through $t^{\prime }+R(t^{\prime })/c=t$ they depended implicitly
on $x,y,z,t$. However, later (in calculating the formula (63.7)) they used
the explicit dependence of $R$ on the space coordinates of the
observation point too. 
Jackson~\cite{jackson} agrees with~\cite{landau} that one should find ``a contribution to the
spatial partial derivative for fixed time $t$ from explicit spatial
coordinate dependence (of the observation point).''} 
\v{S}kovrlj and Ivezi\'{c}~\cite{ivezic} call this partial derivative as `{\it complete} partial
derivative'; Chubykalo and Vlayev,
as `{\it total} \linebreak derivative with respect to
a given variable'.  The terminology suggested by Brownstein~\cite{brown} is
`the {\it whole}-partial derivative'.

\section{Example 1.}

Let us study the case when we deal with explicite and implicite dependencies 
$f ({\bf p}, E ({\bf p}))$. It is well known that the energy in
relativism is related to the 3-momentum as $E=\pm \sqrt{{\bf p}^2 +m^2}$%
; the unit system $c=\hbar=1$ is used. In other words, we must choose the
3-dimensional mass hyperboloid in the Minkowski space, and the energy is 
{\it not} an independent quantity anymore. Let us calculate the commutator
of the whole-partial derivatives $\hat\partial /\hat\partial E$ and $\hat\partial / 
\hat\partial p_i$.
In order to make distinction between differentiating the explicit function
and that which contains both explicit and implicit dependencies, the `whole
partial derivative' may be denoted as $\hat\partial$.
In the general case one has 
\begin{equation}
{\frac{\hat\partial f ({\bf p}, E({\bf p})) }{\hat\partial p_i}} \equiv {%
\frac{\partial f ({\bf p}, E({\bf p})) }{\partial p_i}} + {\frac{\partial f (%
{\bf p}, E({\bf p})) }{\partial E}} {\frac{\partial E}{\partial p_i}}\, .
\end{equation}
Applying this rule, we find surprisingly 
\begin{eqnarray}
&&[{\frac{\hat\partial }{\hat\partial p_i}},{\frac{\hat\partial }{\hat
\partial E}}]_- f ({\bf p},E ({\bf p})) = {\frac{\hat\partial }{\hat\partial
p_i}} {\frac{\partial f }{\partial E}} -{\frac{\partial }{\partial E}} ({
\frac{\partial f}{\partial p_i}} +{\frac{\partial f}{\partial E}}{\frac{%
\partial E}{\partial p_i}}) =  \nonumber \\
&=& {\frac{\partial^2 f }{\partial E\partial p_i}} + {\frac{\partial^2 f}{%
\partial E^2}}{\frac{\partial E}{\partial p_i}} - {\frac{\partial^2 f }{%
\partial p_i \partial E}} - {\frac{\partial^2 f}{\partial E^2}}{\frac{%
\partial E}{\partial p_i}}- {\frac{\partial f }{\partial E}} {\frac{\partial%
}{\partial E}}({\frac{\partial E}{\partial p_i}}).  \label{com}
\end{eqnarray}
So, if $E=\pm \sqrt{m^2+{\bf p}^2}$ 
and one uses the generally-accepted 
representation form of $\partial E/\partial p_i
=  p_i/E$,
one has that the expression (\ref{com})
appears to be equal to $(p_i/E^2) {\frac{\partial f({\bf p}, E ({\bf p}))}{%
\partial E}}$. Within the choice of the normalization the coefficient may be related to 
the longitudinal electric field in the helicity basis.\footnote{The electric/magnetic
fields can be derived from the 4-potentials which have been presented in~ 
\cite{hb}.} Next, the commutator is
\begin{equation}
[{\frac{\hat\partial}{\hat\partial p_i}}, {\frac{\hat\partial}{\hat\partial
p_j}}]_- f ({\bf p},E ({\bf p})) = {\frac{1}{\vert E\vert^3}} {\frac{%
\partial f({\bf p}, E ({\bf p}))}{\partial E}} [p_i, p_j]_-\,.
\end{equation}
This should also not  be zero according to
Feynman and \linebreak Dyson~\cite{FD}. They postulated that the velocity (or, of course, the 3-momentum)
commutator is equal to $[p_i,p_j]\sim i\hbar\epsilon_{ijk} B^k$, i.e., to
the magnetic field.
In fact, if we put in the corespondence to the momenta their
quantum-mechanical operators (of course, with the appropriate clarification $%
\partial \rightarrow \hat\partial$), we obtain again that, in general, the
derivatives do {\it not} commute $[{\frac{\hat\partial}{\hat\partial x_\mu}}%
, {\frac{\hat\partial}{\hat\partial x_\nu}}]_- \neq 0$.

Furthermore, since the energy derivative corresponds to the operator of time
and the $i$-component momentum derivative, to $\hat x_i$, we put forward the
following anzatz in the momentum representation: 
\begin{equation}
[\hat x^\mu, \hat x^\nu]_- = \omega ({\bf p}, E({\bf p})) \,
F^{\mu\nu}_{\vert\vert} ({\bf p}) {\frac{\partial }{\partial E}}\,,
\end{equation}
with some weight function $\omega$ being different for different \linebreak choices of
the antisymmetric tensor spin basis. The physical dimension of $x^\mu$ is $[energy]^{-1}$
in this unit system; $F^{\mu\nu}_{\vert\vert} ({\bf p})$  has the dimension $[energy]^{0}$,
if we assume the mass shell condition in the definition of the field operators $\delta (p^2 -m^2)$,
see~\cite{VVD1}. Therefore, the weight function should have the dimension $[energy]^{-1}$.
The commutator $[\hat x^\mu, \hat p^\nu ]$ has the same form as in the textbook nonrelativistic quantum mechanics 
within the presented model.

In the modern literature, the idea of the broken Lorentz invariance by this
method concurs with the idea of the {\it fundamental length}, first
introduced by V. G. Kadyshevsky~\cite{kadysh} on the basis of old papers by
M. Markov. Both ideas and corresponding theories are extensively discussed.
In my opinion, the main question is: what is the space
scale, when the relativity theory becomes incorrect.

\section{Example 2.}

In the previous Section (see also the paper~\cite{VVDPE}) we found some intrinsic contradictions related to the mathematical 
\linebreak foundations of modern physics.
It is well known that the partial derivatives commute in the Minkowski space (as well as in the 4-dimensional momentum space).
However, if we consider that energy is an implicit function of the 3-momenta and mass (thus, approaching the mass hyperboloid formalism,
$E^2 - {\bf p}^2 c^2 = m^2 c^4$)
then we may be interested in the commutators of  the whole-partial derivatives~\cite{brown} instead. The whole-partial derivatives do not commute, as you see above. If they are associated with the corresponding physical operators, we would have the uncertainty relations in this case. This is an intrinsic contradiction of the SRT. While we start from the same postulates, on using two different ways of reasoning we arrive at the two different physical conclusions.

In this Section I would like to ask another question. \linebreak  Sometimes, when calculating dynamical invariants (and other physical quantities in quantum field theory), and when studying the corresponding massless limits we need to calculate iterated limits. We may encounter a rare situation when two iterated limits are not equal each other in physics. See, for example, Ref.~\cite{VVD1}.
We were puzzled calculating the iterated limits of the aggregate $\frac{E^2 - {\bf p}^2}{m^2}$ (or the inverse one, $\frac{m^2}{E^2 -{\bf p}^2}$, \,\, $c=\hbar=1$): 
\begin{eqnarray}
\lim\limits_{m\to 0} \lim\limits_{E\to \pm\sqrt{{\bf p}^2 +m^2}} (\frac{m^2}{E^2 -{\bf p}^2}) &=& 1\,,\\
\lim\limits_{E\to \pm\sqrt{{\bf p}^2 +m^2}} \lim\limits_{m\to 0} (\frac{m^2}{E^2 -{\bf p}^2}) &=& 0\,.
\end{eqnarray}
Similar mathematical examples are  presented in~\cite{wiki}. 
Physics should have well-defined dynamical invariants. Which iterated limit should be applied in the study of massless limits?
The question of the iterated limits was studied in~\cite{Ilyin}. However, the answers
leave room for misunderstandings and contradictions with the experiments.
One can say: ``The two limits are of very different sorts: the limit of 
$E\rightarrow\pm \sqrt{{\bf p}^2 + m^2}$ is a limit 
that subsumes the statement under the theory of Special Relativity. Such limits should be done first."
However, cases exist when the limit $E\rightarrow\pm \sqrt{{\bf p}^2 + m^2}$ cannot be applied
(or its application leads to the loss of the information). For example, we have for the causal 
Green's function used in the scalar field theory and in the $m\rightarrow 0$ quantum electrodynamics (QED), Ref.~\cite{Green}:
\begin{eqnarray}
&&D^c (x) = \frac{1}{(2\pi)^4} \int d^4p \frac{e^{-ip\cdot x}}{m^2 -p^2-i\epsilon} = \\
&=&\frac{1}{4\pi}\delta (\lambda)
- \frac{m}{8\pi \sqrt{\lambda}}\theta (\lambda) [J_1 (m\sqrt{\lambda}) -iN_1 (m\sqrt{\lambda})]+\nonumber\\
&+&
\frac{im}{4\pi^2 \sqrt{-\lambda}} \theta (-\lambda) K_1 (m\sqrt{-\lambda}),\nonumber
\end{eqnarray}
$\lambda = (x^0)^2 -{\bf x}^2$; $J_1, N_1, K_1$ are the Bessel functions of the first order. The application of $E\rightarrow\pm \sqrt{{\bf p}^2 + m^2}-i\delta$ results in non-existence of the integral.
Meanwhile, the massless limit is made in the integrand in the Feynman gauge with no problems.
Please remember that integrals are also the limits of the Riemann integral sums. The $m\rightarrow 0$
limits are made first sometimes.

Next, the application of the mass shell condition in the Weinberg-Tucker-Hammer $2(2S+1)$-formalism leads to the fact that we would not be able to write
the dynamical equation in the covariant form $[\gamma^{\mu\nu}\partial_\mu \partial_\nu -m^2]\Psi_{(6)} (x)=0$. Apart, the information 
about the propagation of longitudinal modes would be lost (cf. formulas (19,20,27,28) of the first paper~\cite{VVD1}). Moreover, the Weinberg equation and the mapping of the Tucker-Hammer
equation to the antisymmetric tensor formalism have different physical contents on the interaction level~\cite{WEIN,TH}.\footnote{
I take this opportunity 
to note that problems (frequently forgotten) may also exist with the direct application of $m\rightarrow 0$ 
in relativistic  quantum 
equations. The case is: when the solutions are constructed on using the relativistic 
boosts in the momentum space the mass may appear 
in the denominator, $\sim 1/m^n$, which cancels the mass terms of the equation giving 
the non-zero corresponding result.}

Next, if we would always apply the mass shell condition first then we come to the derivative paradox of the previous 
Section. 
Finally, the condition $E^2 -{\bf p}^2 = m^2$ does not always imply the generally-accepted special relativity only. 
For instance,
see the Kapuscik work, Ref.~\cite{Kapuscik}, who showed that similar expressions for energy and 
momentum exist for particles with $V > c$ and $m_\infty \in \Re e$.  

Meanwhile, the case $m=0$ appears to be equivalent to the light cone condition $r = ct$, which can be taken
even without the mass shell condition to study the theories extending the special relativity. Not everybody realizes that
it can be used to deduce  the Lorentz transformations between two different reference frames. Just
take squares and use the lineality: $r_1^2 -c^2 t_1^2 =0 = r_2^2 -c^2 t_2^2$. Hence, in $d=1+1$
we have
$x_2 = \gamma (x_1 - v t_1)\,,\quad 
t_2 = \alpha (t_1 - \frac{\beta}{c} x_1)$\,  
with $\alpha=\gamma=1/\sqrt{1-\frac{v^2}{c^2}}$, the Lorentz factor; $\beta =v/c$.

\section{Example 3.}

The problem of explaining mass splitting of leptons ($e,\mu,\tau$) and quarks has a long history. See, for instance,  a method suggested in 
Refs.~\cite{Barut}, and some new insights in~\cite{DVOAACA}. 
Non-commutativity~\cite{snyder} also exhibits interesting peculiarities in
the Dirac case. 
Recently, we analyzed the Sakurai-van der Waerden method of deriving the Dirac
(and higher-spin) equation~\cite{Dvoh}. We can start from
\begin{equation}
(E I^{(2)}-{\bm \sigma}\cdot {\bf p}) (E I^{(2)}+ {\bm\sigma}\cdot
{\bf p} ) \Psi_{(2)} = m^2 \Psi_{(2)} \,,
\end{equation}
or
\begin{eqnarray}
&&(E I^{(4)}+{\bm \alpha}\cdot {\bf p} +m\beta) (E I^{(4)}-{\bm\alpha}\cdot
{\bf p} -m\beta ) \Psi_{(4)} =0\,.\nonumber\\
&&\label{f4}
\end{eqnarray}
$E$ and ${\bf p}$ form the Lorentz 4-momentum.
Obviously, the inverse operators of the Dirac operators exist  in the non- commutative case.
As in the original Dirac work, we have
$\beta^2 = 1\,,\quad
\alpha^i \beta +\beta \alpha^i =0\,,\quad
\alpha^i \alpha^j +\alpha^j \alpha^i =2\delta^{ij}$ \,.
We also postulate non-commutativity relations for the components of 4-momenta:
\begin{equation}
[E, {\bf p}^i]_- = \Theta^{0i} = \theta^i\,,
\end{equation}
as usual. Therefore the equation (\ref{f4}) will {\it not} lead
to the well-known equation $E^2 -{\bf p}^2 = m^2$. Instead, we have
\begin{eqnarray}
&&\hspace{-5mm}\left \{ E^2 - E ({\bm \alpha} \cdot {\bf p})
+({\bm \alpha} \cdot {\bf p}) E - {\bf p}^2 - m^2 - i ({\bm\sigma}\otimes I_{(2)})
[{\bf p}\times {\bf p}] \right \} \Psi_{(4)}\nonumber\\
&&=0\,.
\end{eqnarray}
For the sake of simplicity, we may assume the last term to be zero. Thus, we arrive at
\begin{equation}
\left \{ E^2 - {\bf p}^2 - m^2 -  ({\bm \alpha}\cdot {\bm \theta})
\right \} \Psi_{(4)} = 0\,.
\end{equation} 
We can  apply the unitary transformation. It is known~\cite{Berg,Dvoe}
that one can\footnote{Some relations for the components ${\bf a}$ must be assumed. Moreover, in our case ${\bm \theta}$ 
must not depend on $E$ and ${\bf p}$. Otherwise, we must take the non-commutativity $[E, {\bf p}^i]_-$ into account again.}
$U_1 ({\bm \sigma}\cdot {\bf a}) U_1^{-1} = \sigma_3 \vert {\bf a} \vert$\,.
For ${\bm \alpha}$ matrices we re-write as
\begin{eqnarray}
{\cal U}_1 ({\bm \alpha}\cdot {\bm \theta}) {\cal U}_1^{-1} = \vert {\bm \theta} \vert
\begin{pmatrix}1&0&0&0\cr
0&-1&0&0\cr
0&0&-1&0\cr
0&0&0&1\end{pmatrix} = \alpha_3 \vert {\bm\theta}\vert\,.
\end{eqnarray}
The explicit form of the $U_1$ matrix is ($a_{r,l}= a_1\pm ia_2$):
\begin{eqnarray}
U_1 &=&\frac{1}{\sqrt{2a (a+a_3)}} \begin{pmatrix}a+a_3&a_l\cr
-a_r&a+a_3\end{pmatrix}  = \\
&=&\frac{1}{\sqrt{2a (a+a_3)}} [ a+a_3 + i\sigma_2 a_1 - i\sigma_1 a_2]\,,\nonumber\\ 
{\cal U}_1 &=&\begin{pmatrix}U_1 &0\cr
0& U_1 \end{pmatrix}\,.
\end{eqnarray}
We now apply the second unitary transformation:
\begin{eqnarray}
{\cal U}_2 \alpha_3 {\cal U}_2^\dagger &=&
\begin{pmatrix}1&0&0&0\cr
0&0&0&1\cr
0&0&1&0\cr
0&1&0&0\end{pmatrix} \alpha_3 \begin{pmatrix}1&0&0&0\cr
0&0&0&1\cr
0&0&1&0\cr
0&1&0&0\end{pmatrix} = \nonumber\\
&=&\begin{pmatrix}1&0&0&0\cr
0&1&0&0\cr
0&0&-1&0\cr
0&0&0&-1\end{pmatrix}\,.
\end{eqnarray}
The final equation is then
\begin{equation}
[E^2 -{\bf p}^2 -m^2 -\gamma^5_{chiral} \vert {\bm \theta}\vert ] \Psi^\prime_{(4)} = 0\,.
\end{equation}
In physical terms this implies mass splitting for a Dirac particle over the non-commutative space, $m_{1,2} =\pm \sqrt{m^2 \pm \theta}$. 
This procedure may be attractive as explanation of mass creation and mass splitting in fermions.

\section{Conclusions.}

We found that the commutator of two derivatives may be {\it not} equal to
zero. As a consequence, for instance, the question arises, if the derivative 
$\hat\partial^2 f/\hat\partial p^\nu\hat\partial p^\mu$ is equal to the
derivative $\hat\partial^2 f/\hat\partial p^\mu\hat\partial p^\nu$ in all
cases?\footnote{%
The same question can be put forward when we have differentiation with
respect to the coordinates too, that may have impact on the correct
calculations of the problem of accelerated charge in classical
electrodynamics.} The presented consideration permits us to provide some
bases for non-commutative field theories and induces us to look for further
development of the classical analysis in order to provide a rigorous
mathematical basis for operations with functions which have both explicit
and implicit dependencies. Several other examples are presented.
Thus, while for physicists everything is obvious in the solutions of the paradoxes, 
this is not so for mathematicians.


\section*{Acknowledgements}
I am grateful to participants of conferences where this idea has been discussed.

%

\begin{flushright}\footnotesize
Submitted on February 18, 2019 / Accepted on Month Day, Year
\end{flushright}

\vspace*{-6pt}
\centerline{\rule{72pt}{0.4pt}}
}

\end{document}